\documentclass[prb,aps,amssymb,showpacs,twocolumn,amsmath,floatfix]{revtex4}
\usepackage{tabularx}
\usepackage{bm}
\usepackage{euscript}
\usepackage{epsfig,psfrag,subfigure}
\usepackage{graphicx}
\usepackage{color}
\usepackage{amsfonts}
\usepackage{exscale}
\usepackage{amsbsy}
\usepackage{hyperref}
\input{epsf}

\def\be{\begin{equation}}
\def\ee{\end{equation}}
\def\ba{\begin{eqnarray}}
\def\ea{\end{eqnarray}}

\begin{document}

\title{Topological properties of ferromagnetic superconductors}
\author{Alfred K. C.  Cheung$^1$, and S. Raghu$^{1,2}$ }
\affiliation{$^1$Department of Physics, Stanford University, Stanford, CA 94305}
\affiliation{$^2$SLAC National Accelerator Laboratory, Menlo Park, CA 94025}
\date{\today}

\begin{abstract}
A variety of heavy fermion superconductors, such as UCoGe, UGe$_2$, and URhGe  exhibit a striking coexistence of bulk ferromagnetism and 
superconductivity.  In these systems, the magnetic moment decreases with pressure, and vanishes at a  ferromagnetic quantum critical point (qcp).  Remarkably, the superconductivity in UCoGe varies smoothly with pressure across the qcp and exists in both the ferromagnetic and paramagnetic regimes.  We argue that in UCoGe, spin-orbit interactions stabilize a time-reversal invariant odd-parity superconductor in the high pressure paramagnetic regime.    Based on a simple phenomenological model, we predict that the transition from the paramagnetic normal state to the phase where superconductivity and ferromagnetism coexist, is a first-order transition.  
\end{abstract}


\maketitle

\section{Introduction}
The B-phase of $^3$He ($^3$He-B) is a time-reversal invariant topological superfluid state: it can be visualized, in an appropriately chosen basis,  as a $p_x + ip_y$ paired state of up-spin fermions, a $p_x - ip_y $ paired state of down spin fermions, and a $p_z$ paired state of fermions with opposite spins\cite{balian1963,anderson1975,volovik2009,vollhardt2013}.  In the simple case where there is a single closed Fermi surface enclosing the origin of momentum-space~\cite{footnote1}, it supports gapless surface states of Majorana fermion zero energy modes\cite{volovik2009}.   A two dimensional analog known as the planar phase can be viewed as a $p_x + ip_y$ paired state of up-spin fermions, and $p_x - ip_y $ paired state of down-spin fermions.  Such a system has zero energy modes corresponding to a Kramers doublet of Majorana fermion edge excitations where opposite spin Majorana fermion modes counterpropagate.  
To date, however, an {\it electronic}  analog of $^3$He-B has not been established in an experimentally realized material.  

\begin{figure}
\includegraphics[width=2.5in]{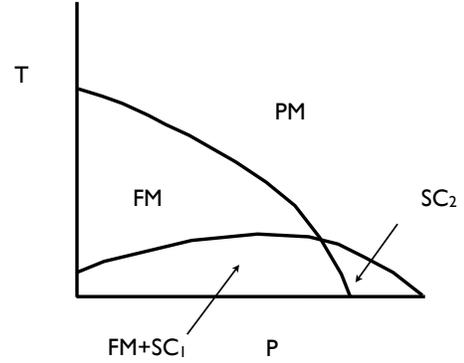}
\caption{Schematic phase diagram of a ferromagnetic superconductor in the presence of strong spin-orbit interactions.  We claim that in the low pressure ferromagnetic (FM) regime, the superconductor SC$_1$ is the B$_2$ phase, whereas in the high pressure paramagnetic (PM) regime, the superconductor SC$_2$ is the B-phase.   }
\label{pd}
\end{figure}

In this paper, we argue that a promising area for observing time-reversal invariant topological superconductivity 
is, ironically, in systems that exhibit a coexistence of ferromagnetism and superconductivity\cite{mineev2011,mineev2012,mineev2014,mineev2015,sau2012,linder2007,linder2008,machida2001}.  Examples of materials in this category include UGe$_2$\cite{saxena2000, huxley2001}, URhGe\cite{aoki2001,levy2005,hardy2005}, and UCoGe~\cite{saxena2000,huy2007,slooten2009,aoki2014,ohta2010}.  Although time-reversal symmetry is certainly broken in the ferromagnetic phase, the Curie temperature can be tuned with pressure, vanishing altogether beyond a quantum critical point $p=p_c$.   We shall focus here on UCoGe, where  
superconductivity occurs  for both ferromagnetic ($p < p_c$), and  paramagnetic ($p>p_c$) regimes\cite{slooten2009,aoki2014}.  A schematic phase diagram of UCoGe is shown in Fig. \ref{pd}.    A particularly striking feature of the phase diagram is the smoothness of $T_c$ as the pressure is tuned past $p_c$.   This feature suggests that the superconductivity for $p>p_c$ and $p < p_c$ must belong to the same irreducible representation.  Motivated by this observation, and considering the effect of the most important energy scales, we will argue that the high pressure regime of UCoGe could be a time-reversal invariant topological superconductor -- an electronic realization of the B-phase.  

Given the complexity of heavy fermion materials, it is difficult to start from a microscopic theory to establish the existence of topological superconductivity in this system.  Instead, we shall take on a more phenomenological approach and take into account the key energy scales.  Specifically, we invoke (i) the proximity to ferromagnetism, (ii) the orthorhombic crystalline symmetry, and (iii) the presence of strong spin-orbit coupling to argue in favor of a ground state that is an electronic analog of of $^3$He-B.  The homogeneous coexistence of ferromagnetism and superconductivity suggests that superconductivity has odd-parity.  The orthorhombic symmetry disfavors the spontaneous breaking of time-reversal symmetry of the superconductivity at high pressures, where ferromagnetism is absent.  Lastly, the strong spin-orbit coupling further favors a B-like phase over a chiral phase.  Based on these  observations, we construct an effective model and predict that the transition from the normal, paramagnetic metal to the coexistence region is a first-order transition.

\section{Phenomenological model}
\begin{figure}
\includegraphics[width=2.5in]{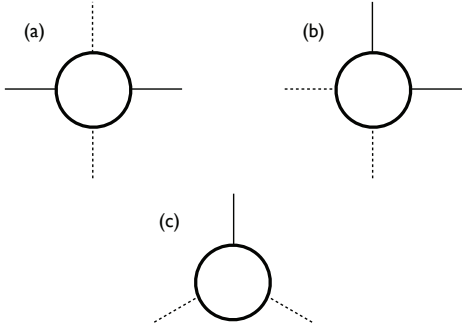}
\caption{Lowest order Feynman diagrams involving the ferromagnetic  (solid external legs) and superconducting (dashed external legs) order parameters.  The solid internal loops represent fermion propagators.  In the mean-field approximation, the diagrams are evaluated at zero energy and momentum.  Diagram (c) is non-zero only for an odd-parity superconductor.   }
\label{diag}
\end{figure} 
The schematic phase diagram of UCoGe is shown in Fig.~\ref{pd}.  
We have referred to the superconductivity that coexists with ferromagnetism as SC$_1$, and have labeled as SC$_2$ the superconductivity that condenses from the paramagnetic normal state.  Generically, there will be a phase transition between SC$_1$ and SC$_2$, since the former breaks time-reversal symmetry whereas the  latter can preserve time-reversal symmetry.

The properties of these materials arise largely from the dynamics of $5f$ electrons of the uranium atom. Let $\Psi_{\bm k, m, \sigma},  (m = -\ell, \cdots , \ell; \sigma = \uparrow, \downarrow)$ 
represent the electron destruction operator for each $f$-orbital ($ \ell = 3$), spin, and momentum state. A low energy effective Hamiltonian that captures all the relevant energy scales has the following form: 
\begin{eqnarray}
H &= &H_0  +  H_{s.o}+H_{ex}+ H_{BCS}. \nonumber \\
\end{eqnarray}
The first term is the kinetic energy in the absence of spin-orbit coupling and magnetism: 
\begin{eqnarray}
H_0&=& \sum_{\bm k , m, \sigma} \epsilon_{\bm k, m}  \Psi^{\dagger}_{\bm k, m, \sigma}    \Psi_{\bm k, m, \sigma}, 
\end{eqnarray}
where $\epsilon_{\bm k,m}  $ are band energies relative to the Fermi level and are taken to be diagonal in both the spins and the orbitals.  
The spin-orbit coupling has the form:
\begin{equation}
H_{s.o}= \frac{\lambda}{2}  \sum_{\bm k} \sum_{m,m',\sigma,\sigma'} \Psi^{\dagger}_{\bm k, m, \sigma} \vec L_{m m'}  \cdot \vec \tau_{\sigma \sigma'} \Psi_{\bm k, m', \sigma'},
\end{equation}
where $L^a_{b,c}$ are  angular momentum matrices in the coordinate representation (for the case of $\ell = 1$, $L^a_{b,c} = -i \epsilon_{abc}$) and $\tau^a$ are Pauli matrices.    The exchange interaction 
is a short-ranged Hubbard-like 
repulsive interaction among electrons in the same orbital that ultimately produces ferromagnetism in the normal state:
\begin{equation}
H_{ex} =  -U \sum_{\bm k, m}  S^z_{\bm k, m}  S^z_{- \bm k, m},
\end{equation}
where $\vec S_{\bm q, m} =  \sum_{\bm k} \sum_{\sigma,\sigma'} \Psi^{\dagger}_{\bm k, \sigma, m}  \vec \tau_{\sigma \sigma'} \Psi_{ \bm k + \bm q, m, \sigma'}$ is the 
Fourier transformed spin density of electrons in orbital $m$.    In the presence of spin-orbit coupling, the exchange interaction has Ising symmetry and the moment lies along a particular axis of the crystal~\cite{ihara2010}.  This interaction can in principle be derived from a knowledge of the microscopic orbitals that are occupied.  In the spirit of our phenomenological treatment, however, we shall make use of the Ising form above.  
Lastly, the superconductivity in this system involves 
quasiparticle states close to the Fermi level and is captured by an effective BCS interaction:
\begin{equation}
H_{BCS} = \sum_{\bm k, \bm k'} \sum_{a , b,\sigma,\sigma'} V_{\bm k, \bm k'}  
\Psi^{\dagger}_{\bm k, a, \sigma} \Psi^{\dagger}_{-\bm k, a, \sigma' } \Psi_{- \bm k', b, \sigma'} \Psi_{ \bm k', b, \sigma}.
\end{equation}
In UCoGe, the Fermi temperature $T_F \sim 40$ K~\cite{hattori2014}, whereas the Curie temperature and maximum superconducting transition temperature are $T_C = 3$ K and $T_{sc} = 0.5$ K respectively~\cite{aoki2014}.  Therefore, it is reasonable to assume the following hierarchy of energy scales: $H_{0}, H_{s.o} > H_{ex} \gg H_{BCS}$.   

\section{Consequences of proximity to ferromagnetism}
The close proximity of superconductivity to ferromagnetism places significant constraints on the possible pairing symmetry in this material, i.e.\ on the form of $V_{\bm k, \bm k'}$.  To obtain insight into this issue, we shall consider the lowest order coupling between the magnetic and superconducting order parameters by decoupling the above interactions via a Hubbard-Stratanovich transformation, and by integrating out fermion fields.  The resulting effective action governs the order parameter fluctuations  in the vicinity of the point where the finite temperature Curie transition and superconducting transition intersect.  The action has the form:
\begin{eqnarray}
S_{eff} =  M^2/U + \int_{\bm k}  \int_{\bm k'} \bar \Delta_{\bm k} V^{-1}_{\bm k, \bm k'} \Delta_{\bm k'} 
+ {\rm Tr} \sum_{n=1}^{\infty} \frac{1}{n} \left(  \mathcal G  \hat O \right)^n ,
\end{eqnarray}   
where $\mathcal  G$  is the Green function in the Nambu basis, $\hat O$ is a matrix involving both the magnetic and superconducting order parameters and depends on the choice of superconducting pairing symmetry, and we have discarded an irrelevant constant involving the logarithm of the normal state Green function.  Keeping the ferromagnetic order parameter $M$ to be Ising-like, with the easy-axis chosen to be along $\hat z$, we consider the possibility of both singlet and triplet superconducting states in the vicinity of the putative tetracritical point, where all order parameters can be taken to be small in a Landau expansion.   For spin triplet states, we employ the standard vector notation for the superconducting order parameter: $\Delta_{\sigma \sigma'}(\bm k) = i \left[ \vec \sigma \cdot \vec d(\bm k) \sigma_y \right]_{\sigma \sigma'}$.  We consider 4 possible scenarios for the superconducting order parameter: 1) singlet pairing, 2) triplet pairing with $ \vec d \parallel \hat z$, 3) triplet pairing with $\vec d \perp \hat z$, and 4) ``non-unitary" triplet superconductivity with $\vert \Delta_{\uparrow\uparrow} \vert \ne \vert \Delta_{\downarrow\downarrow} \vert$.  
  
  The lowest order Feynman diagrams in the expansion of the trace are shown in Fig. \ref{diag}.  The external legs correspond to the order parameter fields (solid lines correspond to ferromagnetism and dashed lines correspond to superconductivity),  and thick solid lines along the loops are fermion propagators.   We shall compute these in the mean-field approximation, setting all momenta and frequencies of external legs to zero.  Diagrams (a), (b) contribute to quartic couplings between the order parameters of the form $c \vert \Delta \vert ^2 M^2$,  and are present both for singlet and triplet superconducting orders.  The coefficient $c$ is  readily expressible in terms of $G_{\pm} = \left( i \omega_n \pm \epsilon(\bm k) \right)^{-1}$, the Matsubara Green functions of the paramagnetic normal state:
  \begin{eqnarray}
  c_{1,2}&=& 2 \int_{\bm k} \sum_{i \omega_n} \left[ G_+^3 G_- + G_-^3 G_+ + G_+^2 G_-^2 \right] \ \ ({\rm cases } \ 1, 2) \nonumber \\
  c_{3,4}&=& 2 \int_{\bm k} \sum_{i \omega_n} \left[ G_+^3 G_- + G_-^3 G_+ - G_+^2 G_-^2 \right] \ \ ({\rm cases } \ 3, 4)  \nonumber \\
  \end{eqnarray}
  The quartic coupling between ferromagnetism and superconductivity is repulsive in all four cases.  It is identical for a singlet superconductor and for a triplet state with $\vec d \parallel \hat z$, and the same is true for a non-unitary triplet superconductor and for a triplet superconductor with $\vec d \perp \hat z$.   It is evident from the expression above that the coefficient $c$ is more repulsive for singlet states and states with $\vec d \parallel \hat z$ than for non-unitary states and for unitary states with $\vec d \perp \hat z$.  The difference in the coefficient $c$ between cases (1,2) and cases (3,4) can be obtained in a standard fashion: one finds that this difference is  $c_{1,2} - c_{3,4} = 7 \beta^2 \rho_0 \zeta(3)/(2 \pi^2)$, where $\beta = 1/T$, $\rho_0$ is the density of states at the Fermi energy, and $\zeta(z)$ is the Riemann-Zeta function.  Thus, the bi-quadratic coupling between ferromagnetism and superconductivity is less repulsive when the $d$-vector is perpendicular to the ferromagnetic moment.  
   
 By contrast, Diagram (c) represents a cubic coupling and is present only for a spin-triplet superconductor with $\vec d \perp \hat z$. It is a coupling of the form $ \gamma M(\vert \Delta_{\uparrow\uparrow} \vert^2 - \vert \Delta_{\downarrow\downarrow} \vert^2)$, where the constant $\gamma$ is: 
 \begin{equation}
 \gamma = \int_{\bm k} \sum_{i \omega_n} \left[ G_-^2(k) G_+(k)-G_+^2(k) G_-(k) \right].
 \end{equation}
 The coefficients above can readily be expressed in terms of derivatives of the density of states at the Fermi level.  However, a qualitative point that ought to be stressed is that while the coefficient $c$ is manifestly positive in all cases, the sign of  $\gamma$ is opposite for the two spin species.  Furthermore, near the transition from the normal, paramagnetic metal to the coexistence phase of ferromagnetism and superconductivity, the cubic coupling plays a more important role.  As we shall see below, it leads to first order transitions.  
   
The main result of this analysis is the intuitive notion that the most likely form of superconductivity in this material is spin triplet superconductivity with $\vec d \perp \hat z$.  For these states, the magnetism does not act as a pair breaker.  
This conclusion ought to be relatively robust and independent of microscopic details, so long as inversion symmetry is preserved.  
  
\section{Effect of crystalline symmetry and spin-orbit coupling}
In this section, we analyze the problem in the vicinity of the superconducting phase transition.  Here, we can obtain robust, model independent conclusions since all allowed gap functions are solutions of the linearized BCS gap equations and their form is dictated largely by the symmetry properties of the normal state.  The relevant symmetry considerations stem from the fact that 1) UCoGe is an orthorhombic material, and 2) spin-orbit coupling is among the largest energy scales, requiring all symmetry transformations to act simultaneously on the the spin and the momentum.  In an orthorhombic system, when the normal state is paramagnetic, the point group symmetries include 180 degree rotations about the crystalline axes, inversion, and reflections about each crystalline axis.  In such a system, there are 4 irreducible representations,  all of which are non-degenerate and correspond to time-reversal invariant phases:
\begin{eqnarray}
A_{1 u}: \vec d(\bm k) &=& \alpha k_x \hat x + \beta k_y \hat y + \gamma k_z \hat z \nonumber \\
B_{1 u}: \vec d(\bm k) &=& \alpha k_y \hat x + \beta k_x \hat y + \gamma k_x k_y k_z \hat z \nonumber \\
B_{2 u}: \vec d(\bm k) &=& \alpha k_z \hat x + \beta k_x k_y k_z \hat y + \gamma k_x \hat z \nonumber \\
B_{3 u}: \vec d(\bm k) &=& \alpha k_x k_y k_z\hat x + \beta k_z \hat y + \gamma k_y  \hat z, \nonumber \\
\end{eqnarray}
where we again describe the order parameter in the standard vector form, via $\Delta_{\sigma \sigma'}(\bm k) = i \left[ \vec \sigma \cdot \vec d(\bm k) \sigma_y \right]_{\sigma \sigma'}$, and the quantities $\alpha,\beta, \gamma$ are real numbers.  The first state corresponds to the Balian-Werthamer state\cite{balian1963} of  $^3$He, in an orthorhombic system.  It is a fully gapped state in both $d=3$ and $d=2$ (in the latter case, $\gamma = 0$).  The remaining 3 solutions possess point nodes on a closed Fermi surface in the vicinity of which the gap vanishes quadratically.  We stress that since symmetry allowed gap functions are non-degenerate irreducible representations, time-reversal symmetry cannot be spontaneously broken in the paramagnetic regime. This rules out the possibility of the A-phase near $T_c$ and the B-phase is the state that is most likely to be realized in the paramagnetic regime.  

Next, we consider the symmetry allowed solutions in the orthorhombic system in the ferromagnetic side.  Since time-reversal is broken by ferromagnetism, the resulting point group symmetries are somewhat different than the paramagnetic case.  Suppose the moments lie along the $\hat z$ direction, the symmetries of the system include 1) 180-degree rotation about the z-axis, 2) reflection about the z-axis, 3) reflections about the x,y axes followed by time-reversal, and 4) 180-degree rotation about the x,y axes followed by time-reversal.  In this case, the simplest allowed gap functions have the form:
\begin{equation}
A_{1 u}: \vec d(\bm k) = (\alpha k_x + i \beta k_y)(\hat x - i \hat y) + (\gamma k_x + i \delta k_y) (\hat x + i \hat y),
\end{equation}
where $\alpha,..., \delta$ are real but their sign is arbitrary.  Physically, this state consists of a superposition of $p_x \pm i p_y$ pairing for both up and down spins.  In two dimensions, such a state is fully gapped whereas in 3 dimensions, it possesses point nodes with Weyl Fermions on a closed Fermi surface.  Thus, it is quite natural to expect that as the moment vanishes, the non-unitary state above undergoes a transition to the B-phase in the paramagnetic side, which also belongs to the $A_{1u}$ irreducible representation.  The transition observed in UCoGe would thus be analogous to the transition between the B phase and B$_2$ phase of $^3$He.

To summarize the analysis of this section, we have taken into account the orthorhombic symmetry and spin-orbit coupling to conclude that the only allowed irreducible representations correspond to B-like states.  The orthorhombic symmetry precludes multi-dimensional irreducible representations that may lead to spontaneous time-reversal symmetry breaking of the high pressure superconductor.  Furthermore, the fact that the superconducting $T_c$ as a function of pressure exhibits smooth behavior suggests that the irreducible representation is likely unchanged as a function of pressure.  This in turn implies that the ferromagnetic superconductor is an analog of the B$_2$ phase, i.e. the B-phase in a Zeeman field.  In the case when the Fermi surface is closed, such a superconductor must necessarily possess point nodes.  While strong coupling feedback effects, which stabilize the $A$ phase in $^3$He, may occur here, they will have to overcome the constraints above imposed by crystalline symmetry -- which are absent in $^3$He.

\section{First-order transition into the coexistence phase}
Based on the considerations above, we now study the generic phase diagram of our system within the framework of  Landau theory. Given the presence of strong spin-orbit coupling and the observation of easy-axis ferromagnetism, we take our magnetization to be an Ising field.
Since the analysis above lead to the conclusion that a superconductor with $\vec d \perp \vec M$ is most favored, we shall take the superconducting order parameter to consist of equal spin pairing along the axis of the magnetization: i.e.\ we take the two components of the order parameter to be $\Delta_{\uparrow \uparrow}$ and $\Delta_{\downarrow \downarrow}$ and neglect opposite spin-pairing components $\Delta_{\uparrow \downarrow}$.   This approximation is sufficient for the present analysis, where we consider phase transitions into the ferromagnetic superconductor.

Therefore, under these simplifications, a Landau theory can be constructed based on three order parameters: $M$, $\Delta_{\uparrow\uparrow}$, and $\Delta_{\downarrow\downarrow}$. In the absence of coupling between the magnetic and superconducting orders, the free energy density takes the form:
\begin{widetext}
\begin{equation}
f_0 = \frac{1}{2}A_M M^2 + B_M M^4 + \frac{1}{2}A_\Delta(|\Delta_{\uparrow\uparrow}|^2+|\Delta_{\downarrow\downarrow}|^2) + B_\Delta(|\Delta_{\uparrow\uparrow}|^2+|\Delta_{\downarrow\downarrow}|^2)^2 + C_\Delta(|\Delta_{\uparrow\uparrow}|^2-|\Delta_{\downarrow\downarrow}|^2)^2.
\label{eq:2}
\end{equation}
\end{widetext}

As we have discussed above, the most relevant coupling between magnetism and superconductivity is of the form:
\begin{equation}
\delta f_1 = \lambda_1 M(|\Delta_{\uparrow\uparrow}|^2-|\Delta_{\downarrow\downarrow}|^2).
\label{eq:3}
\end{equation}
Finally, in the presence of spin-orbit coupling, spin is no longer a conserved quantity. Hence, one might envision a spin non-conserving process in which a spin-up Cooper pair scatters into a spin-down Cooper pair. Such a process could be described by:
\begin{equation}
\delta f_2 = \lambda_2 (\Delta_{\uparrow\uparrow}^{*} \Delta_{\downarrow\downarrow}+c.c.).
\label{eq:4}
\end{equation}

\begin{figure}[t]
\includegraphics[width=\columnwidth]{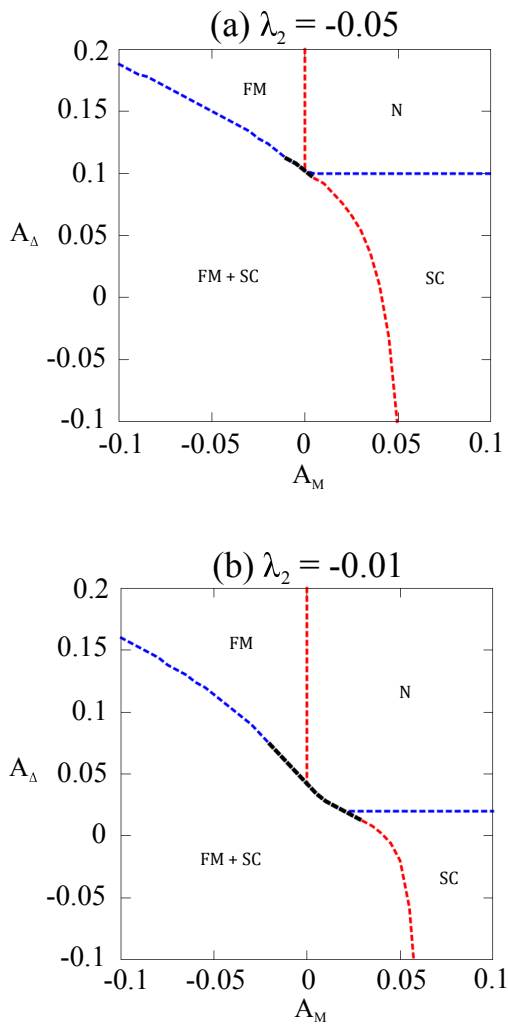}
\caption{Phase diagrams in the $A_\Delta / A_M$ plane constructed by minimizing the free energy $f=f_0 + \delta f_1 + \delta f_2$. The parameters used are $B_M = 1$, $B_\Delta = 1$, $C_\Delta = 2$, $\lambda_1 = 0.5$, (a) $\lambda_2 = -0.05$, and (b) $\lambda_2 = -0.01$. The blue line denotes the superconducting phase transition and the red line denotes the ferromagnetic phase transition. The four phases of UCoGe are reproduced with this model: the normal phase (N), the ferromagnetic phase (FM), the superconducting phase (SC), and the ferromagnetic superconducting phase (FM+SC). The model predicts the generic existence of first order phase transitions (black line) near the boundary between the N and FM+SC phases.}
\label{fig:1}
\end{figure}

We have minimized the free energy $f=f_0 + \delta f_1 + \delta f_2$ and have studied the resulting phase diagrams in the $A_\Delta / A_M$ plane. In Fig.~\ref{fig:1}a, the phase diagram is shown for the set of parameters: $B_M = 1$, $B_\Delta = 1$, $C_\Delta = 2$, $\lambda_1 = 0.5$, and $\lambda_2 = -0.05$. The blue line denotes the superconducting phase transition, the red line denotes the ferromagnetic phase transition, and the thick black line denotes where these transitions become first order. Firstly, note that our minimal phenomenological model is  able to produce the general structure of the phase diagram of UCoGe, namely, the existence of the four phases: a normal phase (N), a purely ferromagnetic phase (FM), a ferromagnetic superconducting phase (FM+SC), and a purely superconducting phase (SC). 

The existence of the SC phase deserves further comment as previous works found that the superconducting order was always accompanied by ferromagnetic order~\cite{shopova2009}. We find that for $C_\Delta=0$, this is indeed the case. The reason for this is apparent by examining the form of $V_1 = \lambda_1 M(|\Delta_{\uparrow\uparrow}|^2-|\Delta_{\downarrow\downarrow}|^2)$. Because $V_1$ is linear in $M$, when $|\Delta_{\uparrow\uparrow}|^2-|\Delta_{\downarrow\downarrow}|^2$ is non-zero, the minimum of free energy must be shifted away from $M=0$. In order to stabilize a purely superconducting phase, a term which favors $|\Delta_{\uparrow\uparrow}|^2=|\Delta_{\downarrow\downarrow}|^2$ must be present. If $|\Delta_{\uparrow\uparrow}|^2=|\Delta_{\downarrow\downarrow}|^2$, then $V_1$ vanishes and the superconducting order can exist with $M=0$. A term of the form $C_\Delta(|\Delta_{\uparrow\uparrow}|^2-|\Delta_{\downarrow\downarrow}|^2)^2$ with $C_\Delta > 0$ provides such an effect. Its inclusion is therefore critical to constructing a proper phenomenological theory for UCoGe.

The existence of the first order transition in the region near the N $\rightarrow$ FM+SC phase transition is found to be a robust feature of our model insensitive to the choice of parameters. This agrees with our discussion of the overall cubic order of $V_1$, and with prior work~\cite{shopova2009}. The location of the first order transition is also as expected -- the transition directly from the N phase to the FM+SC phase is where the magnetic and superconducting order parameters can grow from zero equally. Notably, this property of $V_1$ also induces the nearby FM $\rightarrow$ FM+SC and SC $\rightarrow$ FM+SC phase transitions to be first order. 

Finally, we comment on the effects of varying $\lambda_2$ on the structure of the phase diagram. Shown in Fig.~\ref{fig:1}b is the phase diagram for the same model parameters as in Fig.~\ref{fig:1}a, except with $\lambda_2 = -0.01$. Increasing $\lambda_2$ stabilizes the purely superconducting phase at more positive values of $A_\Delta$ and also decreases the range of the N $\rightarrow$ FM+SC transition. Beyond a certain threshold value for $\lambda_2$, the N $\rightarrow$ FM+SC transition collapses to a point. The case of $\lambda_2 = -0.05$ is already past this threshold. However, the overall structure of the phase diagram is preserved including the existence of the first order transition.  

To summarize, we have presented in this section a simple Landau free energy theory appropriate  for UCoGe. Our minimal model is able to reproduce the general structure of the UCoGe phase diagram and predicts a first order phase transition near the boundary between the normal phase and ferromagnetic superconducting phase. The existence of this first order transition is a necessary consequence of the cubic coupling term between the magnetic and superconducting order parameters which is allowed by time reversal symmetry. As such, we claim its existence to be a robust, generic property of UCoGe independent of any specific microscopic model.

\section{Discussion}
In this paper, we have constructed a phenomenology of the phase diagram of UCoGe, and have argued that the superconductivity in this system is an electronic analog of the B-phase of Helium-3.  Our argument is based on the proximity to ferromagnetism, the orthorhombic crystalline symmetry, and the role of strong spin-orbit coupling.  We highlight that a consequence of such superconducting states is that the apparent ``tetracritical" point in the phase diagram of this material is absent, and the transitions in its vicinity are likely first order transitions.  

To make further progress, a more microscopic treatment is clearly desired.  In this regard it would be interesting to obtain information of the topology of the Fermi surface in this material based on {\it ab initio} calculations.  This information would enable us, in conjunction with our phenomenological arguments above, to state whether Majorana fermion surface states in this material would be topologically protected.  In the future, we would also like to study topological excitations in this system; in the low pressure phase where ferromagnetism and superconductivity coexist, topological excitations are Ising domain walls for ferromagnetism, and ordinary vortices of the superconductor.  The manner in which such domain walls trap fermion zero modes is an interesting question and their experimental signatures remain unclear.  We wish to address these issues in future work.  

\acknowledgments{We thank S.-B. Chung, C. Xu, and especially A. Vishwanath for insightful discussions.  This work was supported in part by DOE Office of Basic Energy Sciences, contract DE- AC02-76SF00515 (SR), the Alfred P. Sloan Foundation (SR), and a NSERC PGS-D Scholarship (AC).  }

\bibliography{ucoge}

\begin{thebibliography}{25}
\expandafter\ifx\csname natexlab\endcsname\relax\def\natexlab#1{#1}\fi
\expandafter\ifx\csname bibnamefont\endcsname\relax
  \def\bibnamefont#1{#1}\fi
\expandafter\ifx\csname bibfnamefont\endcsname\relax
  \def\bibfnamefont#1{#1}\fi
\expandafter\ifx\csname citenamefont\endcsname\relax
  \def\citenamefont#1{#1}\fi
\expandafter\ifx\csname url\endcsname\relax
  \def\url#1{\texttt{#1}}\fi
\expandafter\ifx\csname urlprefix\endcsname\relax\def\urlprefix{URL }\fi
\providecommand{\bibinfo}[2]{#2}
\providecommand{\eprint}[2][]{\url{#2}}

\bibitem[{\citenamefont{Balian and Werthamer}(1963)}]{balian1963}
\bibinfo{author}{\bibfnamefont{R.}~\bibnamefont{Balian}} \bibnamefont{and}
  \bibinfo{author}{\bibfnamefont{N.}~\bibnamefont{Werthamer}},
  \bibinfo{journal}{Phys. Rev.} \textbf{\bibinfo{volume}{131}},
  \bibinfo{pages}{1553} (\bibinfo{year}{1963}).

\bibitem[{\citenamefont{Anderson and Brinkman}(1975)}]{anderson1975}
\bibinfo{author}{\bibfnamefont{P.~W.} \bibnamefont{Anderson}} \bibnamefont{and}
  \bibinfo{author}{\bibfnamefont{W.~F.} \bibnamefont{Brinkman}}, in
  \emph{\bibinfo{booktitle}{15th Scottish Universities Summer School}}, edited
  by \bibinfo{editor}{\bibfnamefont{J.~G.~M.} \bibnamefont{Armitage}}
  \bibnamefont{and} \bibinfo{editor}{\bibfnamefont{I.~E.}
  \bibnamefont{Farquhar}} (\bibinfo{publisher}{Academic Press, New York},
  \bibinfo{year}{1975}).

\bibitem[{\citenamefont{Volovik and Volovik}(2009)}]{volovik2009}
\bibinfo{author}{\bibfnamefont{G.~E.} \bibnamefont{Volovik}} \bibnamefont{and}
  \bibinfo{author}{\bibfnamefont{G.}~\bibnamefont{Volovik}}
  (\bibinfo{year}{2009}).

\bibitem[{\citenamefont{Vollhardt and Wolfle}(2013)}]{vollhardt2013}
\bibinfo{author}{\bibfnamefont{D.}~\bibnamefont{Vollhardt}} \bibnamefont{and}
  \bibinfo{author}{\bibfnamefont{P.}~\bibnamefont{Wolfle}},
  \emph{\bibinfo{title}{The superfluid phases of Helium 3}}
  (\bibinfo{publisher}{Courier Corporation}, \bibinfo{year}{2013}).

\bibitem[{foo()}]{footnote1}
\bibinfo{note}{More generally, protected Majorana fermion surface states occur
  when there are an odd number of closed Fermi surfaces enclosing time-reversal
  invariant crystalline momentum points.}

\bibitem[{\citenamefont{Mineev}(2011)}]{mineev2011}
\bibinfo{author}{\bibfnamefont{V.~P.} \bibnamefont{Mineev}},
  \bibinfo{journal}{Comptes Rendus Physique} \textbf{\bibinfo{volume}{12}},
  \bibinfo{pages}{567} (\bibinfo{year}{2011}).

\bibitem[{\citenamefont{Mineev}(2012)}]{mineev2012}
\bibinfo{author}{\bibfnamefont{V.~P.} \bibnamefont{Mineev}},
  \bibinfo{journal}{Journal of Physics: Conference Series}
  \textbf{\bibinfo{volume}{400}}, \bibinfo{pages}{032053}
  (\bibinfo{year}{2012}),
  \urlprefix\url{http://stacks.iop.org/1742-6596/400/i=3/a=032053}.

\bibitem[{\citenamefont{Mineev}(2014)}]{mineev2014}
\bibinfo{author}{\bibfnamefont{V.~P.} \bibnamefont{Mineev}},
  \bibinfo{journal}{Phys. Rev. B} \textbf{\bibinfo{volume}{90}},
  \bibinfo{pages}{064506} (\bibinfo{year}{2014}),
  \urlprefix\url{http://link.aps.org/doi/10.1103/PhysRevB.90.064506}.

\bibitem[{\citenamefont{Mineev}(2015)}]{mineev2015}
\bibinfo{author}{\bibfnamefont{V.~P.} \bibnamefont{Mineev}},
  \bibinfo{journal}{Phys. Rev. B} \textbf{\bibinfo{volume}{91}},
  \bibinfo{pages}{014506} (\bibinfo{year}{2015}),
  \urlprefix\url{http://link.aps.org/doi/10.1103/PhysRevB.91.014506}.

\bibitem[{\citenamefont{Sau and Tewari}(2012)}]{sau2012}
\bibinfo{author}{\bibfnamefont{J.~D.} \bibnamefont{Sau}} \bibnamefont{and}
  \bibinfo{author}{\bibfnamefont{S.}~\bibnamefont{Tewari}},
  \bibinfo{journal}{Phys. Rev. B} \textbf{\bibinfo{volume}{86}},
  \bibinfo{pages}{104509} (\bibinfo{year}{2012}),
  \urlprefix\url{http://link.aps.org/doi/10.1103/PhysRevB.86.104509}.

\bibitem[{\citenamefont{Linder and Sudb\o{}}(2007)}]{linder2007}
\bibinfo{author}{\bibfnamefont{J.}~\bibnamefont{Linder}} \bibnamefont{and}
  \bibinfo{author}{\bibfnamefont{A.}~\bibnamefont{Sudb\o{}}},
  \bibinfo{journal}{Phys. Rev. B} \textbf{\bibinfo{volume}{76}},
  \bibinfo{pages}{054511} (\bibinfo{year}{2007}),
  \urlprefix\url{http://link.aps.org/doi/10.1103/PhysRevB.76.054511}.

\bibitem[{\citenamefont{Linder et~al.}(2008)\citenamefont{Linder, Sperstad,
  Nevidomskyy, Cuoco, and Sudb\o{}}}]{linder2008}
\bibinfo{author}{\bibfnamefont{J.}~\bibnamefont{Linder}},
  \bibinfo{author}{\bibfnamefont{I.~B.} \bibnamefont{Sperstad}},
  \bibinfo{author}{\bibfnamefont{A.~H.} \bibnamefont{Nevidomskyy}},
  \bibinfo{author}{\bibfnamefont{M.}~\bibnamefont{Cuoco}}, \bibnamefont{and}
  \bibinfo{author}{\bibfnamefont{A.}~\bibnamefont{Sudb\o{}}},
  \bibinfo{journal}{Phys. Rev. B} \textbf{\bibinfo{volume}{77}},
  \bibinfo{pages}{184511} (\bibinfo{year}{2008}),
  \urlprefix\url{http://link.aps.org/doi/10.1103/PhysRevB.77.184511}.

\bibitem[{\citenamefont{Machida and Ohmi}(2001)}]{machida2001}
\bibinfo{author}{\bibfnamefont{K.}~\bibnamefont{Machida}} \bibnamefont{and}
  \bibinfo{author}{\bibfnamefont{T.}~\bibnamefont{Ohmi}},
  \bibinfo{journal}{Phys. Rev. Lett.} \textbf{\bibinfo{volume}{86}},
  \bibinfo{pages}{850} (\bibinfo{year}{2001}),
  \urlprefix\url{http://link.aps.org/doi/10.1103/PhysRevLett.86.850}.

\bibitem[{\citenamefont{Saxena et~al.}(2000)\citenamefont{Saxena, Agarwal,
  Ahilan, Grosche, Haselwimmer, Steiner, Pugh, Walker, Julian, Monthoux
  et~al.}}]{saxena2000}
\bibinfo{author}{\bibfnamefont{S.}~\bibnamefont{Saxena}},
  \bibinfo{author}{\bibfnamefont{P.}~\bibnamefont{Agarwal}},
  \bibinfo{author}{\bibfnamefont{K.}~\bibnamefont{Ahilan}},
  \bibinfo{author}{\bibfnamefont{F.}~\bibnamefont{Grosche}},
  \bibinfo{author}{\bibfnamefont{R.}~\bibnamefont{Haselwimmer}},
  \bibinfo{author}{\bibfnamefont{M.}~\bibnamefont{Steiner}},
  \bibinfo{author}{\bibfnamefont{E.}~\bibnamefont{Pugh}},
  \bibinfo{author}{\bibfnamefont{I.}~\bibnamefont{Walker}},
  \bibinfo{author}{\bibfnamefont{S.}~\bibnamefont{Julian}},
  \bibinfo{author}{\bibfnamefont{P.}~\bibnamefont{Monthoux}},
  \bibnamefont{et~al.}, \bibinfo{journal}{Nature}
  \textbf{\bibinfo{volume}{406}}, \bibinfo{pages}{587} (\bibinfo{year}{2000}).

\bibitem[{\citenamefont{Huxley et~al.}(2001)\citenamefont{Huxley, Sheikin,
  Ressouche, Kernavanois, Braithwaite, Calemczuk, and Flouquet}}]{huxley2001}
\bibinfo{author}{\bibfnamefont{A.}~\bibnamefont{Huxley}},
  \bibinfo{author}{\bibfnamefont{I.}~\bibnamefont{Sheikin}},
  \bibinfo{author}{\bibfnamefont{E.}~\bibnamefont{Ressouche}},
  \bibinfo{author}{\bibfnamefont{N.}~\bibnamefont{Kernavanois}},
  \bibinfo{author}{\bibfnamefont{D.}~\bibnamefont{Braithwaite}},
  \bibinfo{author}{\bibfnamefont{R.}~\bibnamefont{Calemczuk}},
  \bibnamefont{and} \bibinfo{author}{\bibfnamefont{J.}~\bibnamefont{Flouquet}},
  \bibinfo{journal}{Physical Review B} \textbf{\bibinfo{volume}{63}},
  \bibinfo{pages}{144519} (\bibinfo{year}{2001}).

\bibitem[{\citenamefont{Aoki et~al.}(2001)\citenamefont{Aoki, Huxley,
  Ressouche, Braithwaite, Flouquet, Brison, Lhotel, and Paulsen}}]{aoki2001}
\bibinfo{author}{\bibfnamefont{D.}~\bibnamefont{Aoki}},
  \bibinfo{author}{\bibfnamefont{A.}~\bibnamefont{Huxley}},
  \bibinfo{author}{\bibfnamefont{E.}~\bibnamefont{Ressouche}},
  \bibinfo{author}{\bibfnamefont{D.}~\bibnamefont{Braithwaite}},
  \bibinfo{author}{\bibfnamefont{J.}~\bibnamefont{Flouquet}},
  \bibinfo{author}{\bibfnamefont{J.-P.} \bibnamefont{Brison}},
  \bibinfo{author}{\bibfnamefont{E.}~\bibnamefont{Lhotel}}, \bibnamefont{and}
  \bibinfo{author}{\bibfnamefont{C.}~\bibnamefont{Paulsen}},
  \bibinfo{journal}{Nature} \textbf{\bibinfo{volume}{413}},
  \bibinfo{pages}{613} (\bibinfo{year}{2001}).

\bibitem[{\citenamefont{L{\'e}vy et~al.}(2005)\citenamefont{L{\'e}vy, Sheikin,
  Grenier, and Huxley}}]{levy2005}
\bibinfo{author}{\bibfnamefont{F.}~\bibnamefont{L{\'e}vy}},
  \bibinfo{author}{\bibfnamefont{I.}~\bibnamefont{Sheikin}},
  \bibinfo{author}{\bibfnamefont{B.}~\bibnamefont{Grenier}}, \bibnamefont{and}
  \bibinfo{author}{\bibfnamefont{A.~D.} \bibnamefont{Huxley}},
  \bibinfo{journal}{Science} \textbf{\bibinfo{volume}{309}},
  \bibinfo{pages}{1343} (\bibinfo{year}{2005}).

\bibitem[{\citenamefont{Hardy and Huxley}(2005)}]{hardy2005}
\bibinfo{author}{\bibfnamefont{F.}~\bibnamefont{Hardy}} \bibnamefont{and}
  \bibinfo{author}{\bibfnamefont{A.~D.} \bibnamefont{Huxley}},
  \bibinfo{journal}{Phys. Rev. Lett.} \textbf{\bibinfo{volume}{94}},
  \bibinfo{pages}{247006} (\bibinfo{year}{2005}),
  \urlprefix\url{http://link.aps.org/doi/10.1103/PhysRevLett.94.247006}.

\bibitem[{\citenamefont{\textrm{N. T. Huy} \textit{et al.}}(2007)}]{huy2007}
\bibinfo{author}{\bibnamefont{\textrm{N. T. Huy} \textit{et al.}}},
  \bibinfo{journal}{Phys. Rev. Lett.} \textbf{\bibinfo{volume}{99}},
  \bibinfo{pages}{067006} (\bibinfo{year}{2007}).

\bibitem[{\citenamefont{\textrm{E. Slooten} {\it et al.}}(2009)}]{slooten2009}
\bibinfo{author}{\bibnamefont{\textrm{E. Slooten} {\it et al.}}},
  \bibinfo{journal}{Phys. Rev. Lett.} \textbf{\bibinfo{volume}{103}},
  \bibinfo{pages}{097003} (\bibinfo{year}{2009}).

\bibitem[{\citenamefont{Aoki and Flouquet}(2014)}]{aoki2014}
\bibinfo{author}{\bibfnamefont{D.}~\bibnamefont{Aoki}} \bibnamefont{and}
  \bibinfo{author}{\bibfnamefont{J.}~\bibnamefont{Flouquet}},
  \bibinfo{journal}{J. Phys. Soc. Jpn.} \textbf{\bibinfo{volume}{83}},
  \bibinfo{pages}{061011} (\bibinfo{year}{2014}).

\bibitem[{\citenamefont{Ohta et~al.}(2010)\citenamefont{Ohta, Hattori, Ishida,
  Nakai, Osaki, Deguchi, K.~Sato, and Satoh}}]{ohta2010}
\bibinfo{author}{\bibfnamefont{T.}~\bibnamefont{Ohta}},
  \bibinfo{author}{\bibfnamefont{T.}~\bibnamefont{Hattori}},
  \bibinfo{author}{\bibfnamefont{K.}~\bibnamefont{Ishida}},
  \bibinfo{author}{\bibfnamefont{Y.}~\bibnamefont{Nakai}},
  \bibinfo{author}{\bibfnamefont{E.}~\bibnamefont{Osaki}},
  \bibinfo{author}{\bibfnamefont{K.}~\bibnamefont{Deguchi}},
  \bibinfo{author}{\bibfnamefont{N.}~\bibnamefont{K.~Sato}}, \bibnamefont{and}
  \bibinfo{author}{\bibfnamefont{I.}~\bibnamefont{Satoh}}, \bibinfo{journal}{J.
  Phys. Soc. of Jpn.} \textbf{\bibinfo{volume}{79}}, \bibinfo{pages}{023707}
  (\bibinfo{year}{2010}).

\bibitem[{\citenamefont{Ihara et~al.}(2010)\citenamefont{Ihara, Hattori,
  Ishida, Nakai, Osaki, Deguchi, Sato, and Satoh}}]{ihara2010}
\bibinfo{author}{\bibfnamefont{Y.}~\bibnamefont{Ihara}},
  \bibinfo{author}{\bibfnamefont{T.}~\bibnamefont{Hattori}},
  \bibinfo{author}{\bibfnamefont{K.}~\bibnamefont{Ishida}},
  \bibinfo{author}{\bibfnamefont{Y.}~\bibnamefont{Nakai}},
  \bibinfo{author}{\bibfnamefont{E.}~\bibnamefont{Osaki}},
  \bibinfo{author}{\bibfnamefont{K.}~\bibnamefont{Deguchi}},
  \bibinfo{author}{\bibfnamefont{N.}~\bibnamefont{Sato}}, \bibnamefont{and}
  \bibinfo{author}{\bibfnamefont{I.}~\bibnamefont{Satoh}},
  \bibinfo{journal}{Phys. Rev. Lett.} \textbf{\bibinfo{volume}{105}},
  \bibinfo{pages}{206403} (\bibinfo{year}{2010}).

\bibitem[{\citenamefont{Hattori et~al.}(2014)\citenamefont{Hattori, Ihara,
  Karube, Sugimoto, Ishida, Deguchi, Sato, and Yamamura}}]{hattori2014}
\bibinfo{author}{\bibfnamefont{T.}~\bibnamefont{Hattori}},
  \bibinfo{author}{\bibfnamefont{Y.}~\bibnamefont{Ihara}},
  \bibinfo{author}{\bibfnamefont{K.}~\bibnamefont{Karube}},
  \bibinfo{author}{\bibfnamefont{D.}~\bibnamefont{Sugimoto}},
  \bibinfo{author}{\bibfnamefont{K.}~\bibnamefont{Ishida}},
  \bibinfo{author}{\bibfnamefont{K.}~\bibnamefont{Deguchi}},
  \bibinfo{author}{\bibfnamefont{N.~K.} \bibnamefont{Sato}}, \bibnamefont{and}
  \bibinfo{author}{\bibfnamefont{T.}~\bibnamefont{Yamamura}},
  \bibinfo{journal}{J. Phys. Soc. Jpn.} \textbf{\bibinfo{volume}{83}},
  \bibinfo{pages}{061012} (\bibinfo{year}{2014}).

\bibitem[{\citenamefont{Shopova and Uzunov}(2009)}]{shopova2009}
\bibinfo{author}{\bibfnamefont{D.~V.} \bibnamefont{Shopova}} \bibnamefont{and}
  \bibinfo{author}{\bibfnamefont{D.~I.} \bibnamefont{Uzunov}},
  \bibinfo{journal}{Phys. Rev. B} \textbf{\bibinfo{volume}{79}},
  \bibinfo{pages}{064501} (\bibinfo{year}{2009}).

\end{thebibliography}

\end{document}